\documentclass[12pt]{article}

\usepackage{amsmath}
\usepackage{amssymb}
\usepackage{graphicx}
\usepackage{epstopdf}
\usepackage{epsfig}
\usepackage{epsf}
\usepackage{rotating}
\usepackage{hyperref}
\usepackage{float}

\usepackage{caption}
\usepackage{subcaption}

\def\napoli{\textit{ Department of Physics and Astronomy,}\\
\textit{ University of New Mexico, Albuquerque, NM 87131, USA}\\}
\def\Title#1{\begin{center} {\Large #1 } \end{center}}
\def\Author#1{\begin{center}{ \sc #1} \end{center}}
\def\Address#1{\begin{center}{ \it #1} \end{center}}

\newenvironment{Abstract}{\begin{quotation}  }{\end{quotation}}
\newenvironment{Presented}{\begin{quotation} \begin{center} 
             PRESENTED AT\end{center}\bigskip 
      \begin{center}\begin{large}}{\end{large}\end{center} \end{quotation}}

\def\beq{\begin{equation}}
\def\eeq#1{\label{#1}\end{equation}}
\def\eeqn{\end{equation}}

\def\beqa{\begin{eqnarray}}
\def\eeqa#1{\label{#1}\end{eqnarray}}
\def\eeqan{\end{eqnarray}}

\let\bar=\overbar

\def\Dslash{\not{\hbox{\kern-4pt $D$}}}
\def\dslash{\not{\hbox{\kern-2pt $\del$}}}

\def\msb{{\bar{\ssstyle M \kern -1pt S}}}

\begin{document}
\begin{titlepage}

\vfill
\Title{A Selection of Three ATLAS B-Physics Results: A Search Beyond the Standard Model, A Precision Measurement, and the Discovery of a New Heavy Meson}
\vfill
\Author{A.~Taylor\footnote{On behalf of the ATLAS Collaboration}}
\Address{\napoli}
\vfill
\date{}                                           

%
\begin{Abstract}
Three recent results from the ATLAS experiment at the LHC are presented. A new excited heavy meson state is observed through its hadronic transition to the ground state. The production cross-section of $B^+$ mesons is measured as a function of transverse momentum $p_T$ and rapidity \textit{y}. A measurement of the $B_s^0 \rightarrow$ J/$\psi \phi$ decay parameters is reported.

\end{Abstract}
\vfill
\begin{Presented}
DPF 2015\\
The Meeting of the American Physical Society\\
Division of Particles and Fields\\
Ann Arbor, Michigan, August 04--08, 2015\\
\end{Presented}
\vfill
\end{titlepage}

\def\thefootnote{\fnsymbol{footnote}}
\setcounter{footnote}{0}

\section{The ATLAS Detector}
The ATLAS experiment at the LHC uses a general-purpose detector$^{[1]}$ consisting of an inner tracker, a calorimeter and a muon spectrometer. The inner detector (ID) directly surrounds the interaction point; it includes a silicon pixel detector (Pixel), a silicon microstrip detector (SCT) and a transition radiation tracker (TRT), and is embedded in an axial 2 T magnetic field. The ID covers the pseudorapidity range $|\eta| <$ 2.5 and is enclosed by a calorimeter system containing electromagnetic and hadronic sections. The calorimeter is surrounded by a large muon spectrometer (MS) inside an air-core toroidal magnet system that contains a combination of monitored drift tubes (MDTs) and cathode strip chambers (CSCs), designed to provide precise position measurements in the bending plane in the range $|\eta| <$ 2.7. In addition, resistive plate chambers (RPCs) and thin gap chambers (TGCs) with a coarse position resolution but a fast response time are used primarily to trigger muons in the ranges $|\eta| <$ 1.05 and 1.05 $< |\eta| <$ 2.4, respectively. RPCs and TGCs are also used to provide position measurements in the non-bending plane and to improve pattern recognition and track reconstruction.

Momentum measurements in the MS are based on track segments formed in at least two of the three stations of the MDTs and the CSCs. The ATLAS trigger system has three levels: the hardware-based Level-1 trigger and the two-stage High Level Trigger (HLT), comprising the Level-2 trigger and Event Filter (EF). At Level-1, the muon trigger searches for patterns of hits satisfying different transverse momentum $p_T$ thresholds using the RPCs and TGCs. The region-of-interest (RoI) around these Level-1 hit patterns then serves as a seed for the HLT muon reconstruction, in which dedicated algorithms are used to incorporate information from both the MS and the ID, achieving a position and momentum resolution close to that provided by the offline muon reconstruction.

We present here three recent B-Physics measurements with this detector.

\section{Observation of an Excited $B_c^\pm$ Meson State with the ATLAS Detector}
The second S-wave state of the $B_c^\pm$, the $B_c^\pm$(2S), is predicted to have a mass in the range 6835$-$6917 MeV. The $B_c^\pm$(2S) state is reconstructed in the decay to the $B_c^\pm$ meson and two oppositely charged pions, with the $B_c^\pm$ reconstructed through its decay to J/$\psi \pi ^\pm$, J/$\psi \rightarrow \mu ^+ \mu^-$.

This study$^{[2]}$ uses pp collision data collected in the years 2011 ($\surd$s = 7 TeV) and 2012 ($\surd$s = 8 TeV) by the ATLAS experiment. The datasets used correspond to an integrated luminosity of 4.9 fb$^{-1}$ and 19.2 fb$^{-1}$, respectively. To improve the resolution, peaks are sought in the distribution of the variable Q = m($B_c^\pm \pi \pi$) $-$ m($B_c^\pm$) $-$ 2m($\pi^\pm$), where m($B_c^\pm$) is the reconstructed invariant mass of the $B_c^\pm$ candidate, m($B_c^\pm \pi \pi$) is the invariant mass of the $B_c^\pm$ candidate combined with two charged pion candidates, and m($\pi^\pm$) is the charged pion mass. The $B_c^\pm$ candidates are reconstructed through the decay $B_c^\pm \rightarrow$J/$\psi (\mu ^+ \mu ^-) \pi^\pm$.

The reconstruction of the excited state candidates uses the $B_c^\pm$ ground state candidates within $\pm3\sigma$ of the fitted mass value of the corresponding dataset. These candidates are combined as described below with two pion candidate tracks associated with the corresponding primary vertex. The $p_T$ threshold of the pion candidates is 400 MeV. No additional selection requirement is applied to the $B_c^\pm$(2S) pion candidates. The three tracks from the secondary vertex and the two tracks from the primary vertex are refitted simultaneously with the following constraints given by the decay topology: the refitted triplet of the $B_c^\pm$  tracks and the pair of primary vertex pion tracks must intersect in two separate $B_c^\pm$ and $B_c^\pm$(2S) vertices. The invariant mass of the refitted muon tracks is constrained to the J/$\psi$ world average mass, and the combined momentum of the refitted $B_c^\pm$ tracks must point to the $B_c^\pm$(2S) vertex. When multiple $B_c^\pm$(2S) candidates are found in the same event, the one with the best $\chi^2$ value returned by the fitter is kept as an excited state candidate. Wrong-charge combinations ($B_c^\pm \pi^+ \pi^+$ and $B_c^\pm \pi^- \pi^-$) are kept separately for comparison with the combinatorial background shape in the right-charge combinations (oppositely charged pion pairs).

A structure is observed in the mass difference distribution. In order to characterize it, an unbinned maximum likelihood fit to the right-charge combinations is performed. The fit includes a third-order polynomial to model the background and a Gaussian function for the structure. The background shape resulting from the fit is verified to be consistent with the wrong-charge combinations (which are not used to constrain the model in the right-charge fit) by fitting the same shape to them, with the normalization as the only free parameter.

The fit finds the peak at a mass difference (Q) value of 288.2 $\pm$ 5.1 MeV in the 7 TeV data and 288.4 $\pm$ 4.8 MeV in the 8 TeV data. The fit yields 22 $\pm$ 6 signal events in the 7 TeV data and 35 $\pm$ 13 events in the 8 TeV data. The Gaussian width of the structure is found to be 18.2 $\pm$ 3.8 MeV in the 7 TeV data and 17.6 $\pm$ 4.0 MeV in the 8 TeV data. Sources of systematic uncertainty include the systematic uncertainty on the mass of the ground state, the systematic uncertainties in the fit of the mass difference distribution itself, alternative models for the signal and the background parameterizations, results from the study of the mass bias in the selection of the candidate with the best $\chi^2$ of the vertex fit, and parameters of the generation taken from the fit with their uncertainties.

The new state is observed at Q = 288.3 $\pm$ 3.5 $\pm$ 4.1 MeV (calculated as the error weighted mean of the 7 TeV and 8 TeV mass values) corresponding to a mass of 6842 $\pm$ 4 $\pm$ 5 MeV, where the first error is statistical and the second is systematic. The significance of the observation is 5.2$\sigma$ with the “look elsewhere effect” taken into account, and the local significance is 5.4$\sigma$. Within the uncertainties, the mass of the resonance corresponding to the observed structure is consistent with the predicted mass of the $B_c^\pm$(2S) state.

\section{Measurement of the differential cross-section of $B^+$ meson production in pp collisions at $\surd$s = 7 TeV at ATLAS}
Measurements of the b-hadron production cross-section in proton-proton collisions at the Large Hadron Collider (LHC) provide tests of QCD calculations for heavy-quark production at higher center-of-mass energies and in wider transverse momentum ($p_T$) and rapidity (\textit{y}) ranges than previously.

A measurement$^{[3]}$ has been made of the $B^+$ production cross-section using the decay channel $B^+ \rightarrow$J/$\psi K^+ \rightarrow \mu^+ \mu^- K^+$ in pp collisions at $\surd$s = 7 TeV, as a function of $B^+$ transverse momentum and rapidity. The results are reported for $B^+$ meson production, but are derived from both charged states, under the assumption that in the phase space accessible by this measurement the $B^+$ and $B^-$ production cross-sections are equal. This assumption is in agreement with the predictions of NLO Monte Carlo generators and is also valid within the precision of the measurement. The collected data correspond to an integrated luminosity of 2.4 fb$^{-1}$ with an uncertainty of 1.8\%.

To compare the cross-section measurements with theoretical predictions, NLO QCD calculations matched with a leading-logarithmic parton shower MC simulation are used. Predictions for $b\bar{b}$ production are evaluated with two packages: POWHEG-HVQ (POWHEG-BOX 1.0)$^{[5],[6]}$ and MC@NLO 4.01$^{[7],[8]}$. The b-quark production cross-section is also calculated in the FONLL theoretical framework$^{[9],[10]}$, permitting direct comparison with the data assuming the world average of the hadronization fraction $f_{\bar{b} \rightarrow B^+}$ = 0.401 $\pm$ 0.008.

The integrated $B^+$ production cross-section in the kinematic range 9 GeV $< p_T <$ 120 GeV and $|y| <$ 2.25 is:
$$\sigma (pp \rightarrow B^+ X) = 10.6 \pm 0.3 (stat.) \pm 0.7 (syst.) \pm 0.2 (lumi.) \pm 0.4 (BR)  \mu b.$$
The FONLL prediction, with its theoretical uncertainty from the renormalization and factorization scale and the b-quark mass, is:
$$\sigma (pp \rightarrow bX)*f_{\bar{b} \rightarrow B^+} = 8.6_{-1.9}^{+3.0} (scale) \pm 0.6 (m_b)  \mu b,$$
where $f_{\bar{b} \rightarrow B^+}$ = (40.1 $\pm$ 0.8)\% is the world-average value for the hadronization fraction. The corresponding predictions of POWHEG and MC@NLO are 9.4 $\mu b$ and 8.8 $\mu b$, respectively, with theoretical uncertainties similar to those of the FONLL prediction.

The cross-section was measured as a function of transverse momentum and rapidity. The next-to-leading-order QCD calculation is compatible with the measured differential cross-section. Within uncertainties, POWHEG+PYTHIA$^{[11]}$ is in agreement with the measured integrated cross-sections and with the dependence on $p_T$ and \textit{y}. At low $|y|$, MC@NLO+HERWIG$^{[12]}$ predicts a lower production cross-section and a softer $p_T$ spectrum than the one observed in data, while for $|y| >$ 1 the predicted $p_T$ spectrum becomes harder than observed in data. The FONLL calculation is in good agreement with the measured differential cross-section d$\sigma$/d$p_T$, within the theoretical uncertainty.

\section{Flavor tagged time dependent angular analysis of the $B_s^0 \rightarrow$ J/$\psi \phi$ decay and extraction of $\Delta \Gamma_s$ and the weak phase $\phi_s$ in ATLAS}
New phenomena beyond the predictions of the Standard Model (SM) may alter CP violation in B-decays. A channel that is expected to be sensitive to new physics contributions is the decay $B_s^0 \rightarrow$J/$\psi \phi$. CP violation in the $B_s^0 \rightarrow$J/$\psi \phi$ decay occurs due to interference between direct decays and decays with $B_s^0-\bar{B}_s^0$ mixing. The CP violating phase $\phi_s$ is defined as the weak phase difference between the $B_s^0-\bar{B}_s^0$ mixing amplitude and the $b \rightarrow c\bar{c}s$ decay amplitude. In the SM the phase $\phi_s$ is small and can be related to CKM quark mixing matrix elements via the relation $\phi_s \approx -2\beta_s$, with $\beta_s=$ arg$[-(V_{ts} V_{tb}^*)⁄(V_{cs} V_{cb}^* )]$; a value of $\phi_s \approx -2\beta_s=-0.037 \pm 0.002$ rad is predicted in the SM. Another physical quantity involved in $B_s^0-\bar{B}_s^0$ mixing is the width difference $\Delta\Gamma_s=\Gamma_L-\Gamma_H$, which is predicted to be $\Delta\Gamma_s = 0.087 \pm 0.021$ ps$^{-1}$ $^{[13],[14]}$. Physics beyond the SM is not expected to affect $\Delta\Gamma_s$ as significantly as $\phi_s$. Extracting $\Delta\Gamma_s$ from data is nevertheless useful as it allows theoretical predictions to be tested.

An update to the previous measurement$^{[15]}$, with the addition of flavor tagging, is presented$^{[4]}$. Flavor tagging significantly reduces the uncertainty of the measured value of $\phi_s$ while also allowing a measurement of one of the strong phases. Previous measurements of these quantities have been reported by the D0$^{[16]}$, CDF$^{[17]}$ and LHCb$^{[18]}$ collaborations. The analysis presented here uses 4.9 fb$^{-1}$ of LHC pp data at $\surd$s = 7 TeV collected by the ATLAS detector in 2011.

The determination of the initial flavor of neutral B-mesons can be inferred using information from the B-meson that is typically produced from the other b-quark in the event. This is referred to as opposite-side tagging (OST). To study and calibrate the OST methods, events containing the decays of $B^\pm \rightarrow$J/$\psi K^\pm$ can be used, where the flavor of the B-meson at production is provided by the kaon charge.

Several parameters describing the $B_s^0$ meson system are measured. These include the mean $B_s^0$ lifetime 1/$\Gamma_s$, the decay width difference $\Delta\Gamma_s$ between the heavy and light mass eigenstates, the transversity amplitudes $|A_0 (0)|$, $|A_{||} (0)|$, and $|A_S|$, and the corresponding strong phases $\delta_{||}$, $\delta_\perp$, and $\delta_S$.

An unbinned maximum likelihood fit is performed on the selected events to extract the parameters of the $B_s^0 \rightarrow$J/$\psi (\mu^+ \mu^-) \phi (K^+ K^-)$ decay. The full simultaneous maximum likelihood fit contains 25 free parameters. These include the nine physics parameters: $\Delta\Gamma_s,\phi_s,\Gamma_s,|A_0 (0)|^2,|A_{||} (0)|^2,\delta_{||},\delta_\perp,|A_S|^2$, and $\delta_S$. The other parameters in the likelihood function are the $B_s^0$ signal fraction $f_s$, the parameters describing the J/$\psi \phi$ mass distribution, the parameters describing the $B_s^0$ meson decay time plus angular distributions of background events, the parameters used to describe the estimated decay time uncertainty distributions for signal and background events, and scale factors between the estimated decay time and mass uncertainties and their true uncertainties. The number of signal $B_s^0$ meson candidates extracted from the fits is 22670 $\pm$ 150. 

The PDF describing the $B_s^0 \rightarrow$ J/$\psi \phi$ decay is invariant under the following simultaneous transformations:
$$[ \phi_s,\Delta\Gamma_s,\delta_\perp,\delta_{||} ] \rightarrow [ \pi-\phi_s,-\Delta\Gamma_s,\pi-\delta_\perp,2\pi-\delta_{||} ].$$
$\Delta\Gamma_s$ has been determined to be positive. Therefore there is a unique solution and only the case $\Delta\Gamma_s >$ 0 is considered. Uncertainties on individual parameters were studied in detail in likelihood scans.

The likelihood behavior of $\delta_\perp$ appears Gaussian and therefore it is reasonable to quote $\delta_\perp = 3.89  \pm 0.47 (stat)$ rad. For $\delta_\perp - \delta_S$ the scan shows a minimum close to $\pi$; however, it is insensitive over the rest of the scan at the level of 2.1$\sigma$. Therefore the measured value of the difference $\delta_\perp - \delta_S$ is only given as 1$\sigma$ confidence interval [3.02, 3.25] rad. For the strong phase $\delta_{||}$ the central fit value is close to $\pi$ $(3.14 \pm 0.10)$ and the 1D likelihood scan shows normal Gaussian behavior around this minimum. However, the systematic pull plot based on 2400 pseudo-experiments fits reveals a double-Gaussian shape with 68\% of the results included in the interval [2.92, 3.35] rad and so we quote the result in the form of a 68\% C.L. interval $\delta_{||}\in$ [2.92, 3.35]  rad (statistical only).

Each is consistent with its respective world average. Likelihood contours in the $\phi_s-\Delta\Gamma_s$ plane are also provided. The fraction $|A_S (0)|^2$, the signal contribution from $B_s^0 \rightarrow$J/$\psi K^+ K^-$ and $B_s^0 \rightarrow$J/$\psi f_0$ decays, is measured to be consistent with zero, at $0.024 \pm 0.014 (stat.) \pm 0.028 (syst.)$.

The results are:

\begin{center}
$\phi_s = 0.12 \pm 0.25 (stat.) \pm 0.05 (syst.)$ rad

$\Delta\Gamma_s = 0.053 \pm 0.021 (stat.) \pm 0.010 (syst.) ps^{-1}$

$\Gamma_s = 0.677 \pm 0.007 (stat.) \pm 0.004 (syst.) ps^{-1}$

$|A_{(||)} (0)|^2 = 0.220 \pm 0.008 (stat.) \pm 0.009 (syst.)$

$|A_0 (0)|^2 = 0.529 \pm 0.006 (stat.) \pm 0.012 (syst.)$

$\delta_\perp = 3.89 \pm 0.47 (stat.) \pm 0.11 (syst.)$ rad
\end{center}

The values are consistent with those obtained in our untagged analysis and significantly reduce the overall uncertainty on $\phi_s$. These results are consistent with the values predicted in the Standard Model.



%
%
%


\clearpage
\begin{thebibliography}{99}
	
	
	\bibitem{ATLAS}
	ATLAS Collaboration, \textit{The ATLAS Experiment at the CERN Large Hadron Collider}, JINST 3 (2008) S08003.
	
	
	\bibitem{ExcitedBc}
	ATLAS Collaboration. \textit{Observation of an Excited $B^±_c$ Meson State with the ATLAS Detector}, Phys. Rev. Lett. 113:212004 (2014).
	
	
	\bibitem{B+}
	ATLAS Collaboration, \textit{Measurement of the differential cross-section of $B^+$ meson production in pp collisions at $\surd$s = 7 TeV at ATLAS}, JHEP 10 (2013) 042.
	
	
	\bibitem{BSM}
	ATLAS Collaboration, \textit{Flavor tagged time dependent angular analysis of the $B_s^0 \rightarrow$ J/$\psi \phi$ decay and extraction of $\Delta \Gamma_s$ and the weak phase $\phi_s$ in ATLAS}, Phys. Rev. D90 (2014), no. 5, 052007.
	
	
	\bibitem{POWHEG1}
	P. Nason, \textit{A new method for combining NLO QCD with shower Monte Carlo algorithms}, JHEP 11 (2004) 040.
	
	
	\bibitem{POWHEG2}
	S. Frixione, P. Nason, and G. Ridolfi, \textit{A positive-weight next-to-leading-order Monte Carlo for heavy flavour hadroproduction}, JHEP 09 (2007) 126.
	
	
	\bibitem{MC@NLO1}
	S. Frixione and B. R. Webber, \textit{Matching NLO QCD computations and parton shower simulations}, JHEP 06 (2002) 029.
	
	
	\bibitem{MC@NLO2}
	S. Frixione, P. Nason, and B. R. Webber, \textit{Matching NLO QCD and parton showers in heavy flavor production}, JHEP 08 (2003) 007.
	
	
	\bibitem{FONLL1}
	P. Nason, S. Dawson, and R. K. Ellis, \textit{The total cross-section for the production of heavy quarks in hadronic collisions}, Nucl. Phys. B 303 (1988) 607.
	
	
	\bibitem{FONLL2}
	P. Nason, S. Dawson, and R. K. Ellis, \textit{The one particle inclusive differential cross section for heavy quark production in hadronic collisions}, Nucl. Phys. B 327 (1989) 49.
	
	
	\bibitem{PYTHIA}
	T. Sjostrand, S. Mrenna, and P. Z. Skands, \textit{Pythia 6.4 Physics and Manual}, JHEP 05 (2006) 026.
	
	
	\bibitem{HERWIG}
	G. Corcella, I. Knowles, G. Marchesini, S. Moretti, K. Odagiri, et al., \textit{HERWIG 6: An event generator for hadron emission reactions with interfering gluons (including supersymmetric processes)}, JHEP 01 (2001) 010.
	
	
	\bibitem{Lenz:2011ti}
	A.~Lenz and U.~Nierste,	\textit{Numerical Updates of Lifetimes and Mixing Parameters of B Mesons}, arXiv:1102.4274 [hep-ph].
	
	
	\bibitem{Lenz:2006hd} 
	A.~Lenz and U.~Nierste,	\textit{Theoretical update of $B_s - \bar{B}_s$ mixing}, JHEP 0706 (2007) 072.
	
	
	\bibitem{Prev.Meas.}
	ATLAS Collaboration, \textit{Time-dependent angular analysis of the decay $B_s^0 \rightarrow$ J/$\psi \phi$ and extraction of $\Delta \Gamma_s$ and the CP-violating weak phase $\phi_s$ by ATLAS}, JHEP 1212 (2012) 072.
	
	
	\bibitem{OtherMeas.1}
	D0 Collaboration, V. M. Abazov et al., \textit{Measurement of the CP-violating phase $\phi_s^{J/\psi \phi}$ using the flavor-tagged decay $B_s^0 \rightarrow$ J/$\psi \phi$ in 8 fb$^{-1}$ of $p\bar{p}$ collisions}, Phys. Rev. D 85 (2012) 032006.
	
	
	\bibitem{OtherMeas.2}
	CDF Collaboration, T. Aaltonen et al., \textit{Measurement of the Bottom-Strange Meson Mixing Phase in the Full CDF Data Set}, Phys. Rev. Lett. 109 (2012) 171802.
	
	
	\bibitem{OtherMeas.3}
	LHCb Collaboration, R. Aaij et al., \textit{Measurement of CP violation and	the $B_s^0$ meson decay width difference with $B_s^0 \rightarrow$ J/$\psi K^+ K^-$ and $B_s^0 \rightarrow$ J/$\psi \pi^+ \pi^-$ decays}, Phys. Rev. D 87 (2013) 112010.
	
\end{thebibliography}
\end{document}